\documentclass[12pt]{article}
\usepackage{epsfig}
\usepackage{graphics}

\begin{document}
 

 
\normalsize
\thispagestyle{empty}
\setcounter{page}{1}

 
\vspace*{0.44truein}
 
\centerline{{\large \bf Hybrid Parallel Computation of Integration in
GRACE}}
\vspace*{0.37truein}
\centerline{Fukuko Yuasa, Tadashi Ishikawa, Setsuya Kawabata}
\vspace*{0.015truein}
\centerline{{KEK, High Energy Accelerator Research Organization, 1-1 OHO}}
\vspace*{0.015truein}
\centerline{{Tsukuba City, Ibaraki 305, Japan}}
\vspace*{10pt}
\centerline{{and}}
\vspace*{10pt}
\centerline{Denis Perret-Gallix}
\vspace*{0.015truein}
\centerline{{LAPP, Laboratoire d'Annecy-le-Vieux de
  Physique des Particules}}
\centerline{{74941 Annecy-le-Vieux CEDEX, France}}
\vspace*{10pt}
\centerline{ {and}}
\vspace*{10pt}
\centerline{Kazuhiro Itakura, Yukihiko Hotta, Motoi Okuda}
\vspace*{0.015truein}
\centerline{ {Fujitsu Corporation, 1-9-3 Nakase Mihama, Chiba 261, Japan}}
\vspace*{0.21truein}

\vspace*{10pt}

\abstract{
With an integrated software package {\tt GRACE}, it is possible to
generate Feynman 
diagrams, calculate the total cross section and generate physics
events automatically. We outline the hybrid method of parallel
computation of the multi-dimensional integration of {\tt GRACE}.
We used {\tt MPI} (Message Passing Interface) as
the parallel library and, to improve the performance we embedded the
mechanism of the dynamic load balancing. The reduction rate of the
practical execution time was studied. 
}{}{}
 
\newpage
 
 
\vspace*{1pt}      

\section{Introduction}          
\vspace*{-0.5pt}
\noindent
The requirements for reducing the practical execution time in {\tt
GRACE} have fostered our interest in parallelization of the 
multi-dimensional integration of {\tt GRACE}\cite{GRACE}. 
In addition, reducing the program size becomes more important to avoid
the cash-miss, which increases 
the practical execution time, as the number of final particles becomes
large. 
These motivations have lead to implement the parallelization of {\tt GRACE}.
\par
In {\tt GRACE}, for the multi-dimensional integration, {\tt
BASES/SPRING}\cite{BASES1,BASES2} is used.
{\tt BASES} is a software package of Monte Carlo integration with an
importance and 
stratified sampling method. For the parallelization of Monte Carlo
integration, it is natural and efficient to distribute sampling points 
to processors. This parallelization is called the Data Parallel.
When the integrand can be decomposed, the Function Parallel is
applicable, that is, each part of the integrand is calculated
in each different processor.
Since 1992, we have investigated independently these two approaches to the
parallel computation of the multi-dimensional integration in {\tt GRACE}. 
\par
Based on above
experiences\cite{parallelbases1,parallelbases2,europvm,pvmgrace}, we 
have developed a new method, a hybrid use
of Data Parallelism and Function Parallelism.
In this method, 
we use {\tt MPI-1}\cite{mpi}\footnote{The first version of {\tt MPI}
standard. It is standardized in May 1994.} as the Message Passing Library which is
standardized and is widely used for developing parallel code in both the
distributed computing environment and MPP (Massively Parallel
Processors) platform.
A computing model adopted is {\tt SPMD} (Single Program Multiple Data)
computing model.  
\par
In this paper, in section 2 the details of the implementation
of hybrid use of Data Parallelism and Function Parallelism
is described. We also present the mechanism 
of the dynamic load balancing in section 3. 
The behavior of performance of the parallel computation is shown
in section 4. 
Section 5 is devoted to a conclusion.

\section{Hybrid Use of Data and Function Parallelism}\label{parallelgrace}
\noindent
In the hybrid method, we firstly distribute 
a group of {\it hypercubes} which are subspaces of the integral volume. In each
{\it hypercube}, a definite number of sampling points are taken. 
The distribution of {\it hypercubes} corresponds to the distribution of
sampling points. 
Secondly, together with distributing sampling points, we distribute 
the calculation of the scattering amplitudes, of which the
integrand consists. 
The scattering amplitude is  given as a sum of each matrix element
corresponding to each Feynman diagram. Therefore the distribution of the
calculation of the scattering amplitudes corresponds  to distribution
Feynman diagrams. 
\par
For the data-transfer among processors,  we use a collective
communication instead of the point-to-point communication. The collective 
communication is defined in a {\it communicator} which
is one of the important concepts in {\tt MPI}.
It defines the communication space used for the communications among
processors. 
\par
\begin{figure}
\begin{center}
\fbox{\epsfig{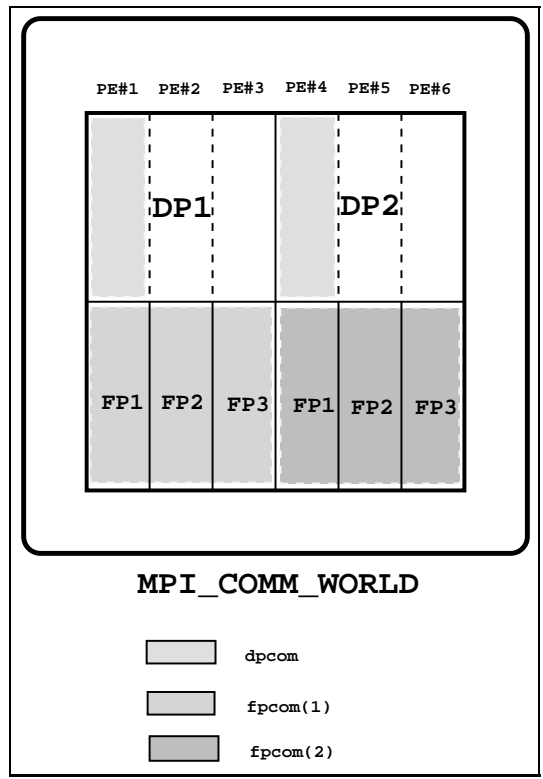}}
\end{center}
\caption{Schematic view of {\it Communicators}. {\tt PE}\#$n$ represents each
parallel processor. {\tt DP1} and {\tt DP2} are groups defined for
distributing {\it hypercubes}. {\tt FP1}, {\tt FP2} and {\tt FP3} are groups
for distributing Feynman diagrams.
{\tt dpcom} and {\tt fpcom}s represent the
{\it communicator} for the Data 
Parallel and for the Function Parallel, respectively.
{\tt MPI\_COMM\_WORLD} is a predefined {\it communicator} 
by {\tt MPI}.
}
\label{fig:hybrid} 
\end{figure}
Fig.~\ref{fig:hybrid} shows the schematic view how the Data
Parallel and the Function Parallel work together from the
point of view of {\it communicators}, where {\it hypercubes}, as an 
example, are 
divided into two groups ({\tt DP1} and {\tt DP2}) and then
calculation of Feynman diagrams into three groups ({\tt
FP1}, {\tt FP2} and {\tt FP3}). 
The {\it communicator} corresponding to the Data Parallelism is indicated
as {\tt dpcom}. On the 
other hand, the {\it communicator} corresponding to the Function
Parallelism is indicated as {\tt fpcom}. 
In {\tt fpcom}, the global sum of the results of the calculation
of the scattering amplitudes from each processor is carried out.  
Successively, in {\tt dpcom} the global sum of several results
needed for estimating the integral is carried out. 
In the figure, note that {\tt dpcom} (hatched in the figure) is
constructed by one of {\tt PE}'s (Processing Elements) in  each group,
{\tt DP1} and {\tt DP2}, to avoid making a global sum repeatedly.
\par
Also note that the total number of parallel processors is
given as the product of the degree of the Data Parallel and that of
the Function Parallel. In this example,  
the number of parallel processors is $ 2 \times  3 = 6$.
\par
Summarizing, the parallel computation of the multi-dimensional
integration is proceeded in the following way: 
\begin{enumerate}
\item{distribute {\it hypercubes} into parallel processors,}
\item{by using sampling points in {\it hypercubes} distributed,
calculate a part 
of integrand (scattering amplitudes) in each parallel processor,}
\item{make a global sum of the scattering amplitudes for each sampling
point in the {\it communicator} {\tt fpcom},}
\item{square the results obtained in (iii) and sum them up in the
group, {\tt DP1} and {\tt DP2},}
\item{make a global sum in the {\it communicator} {\tt dpcom}.} 
\end{enumerate}
The above procedures are iterated until the multi-dimensional
integration converges.

\section{Dynamic Load Balancing}\label{loadbalance}
Once a physics process to be calculated is fixed, {\tt GRACE} generates
the Feynman diagrams according to defined physics model and defined order
of the perturbation. Generated diagrams 
are numbered  
by {\tt GRACE} for convenience. 
Assuming all these diagrams have the
same numbers of vertices and internal lines for
simplicity, the execution time needed to calculate each diagram
is expected to be nearly identical.
On this assumption, we distributed Feynman diagrams into 
parallel processors in the order conventionally numbered by {\tt
GRACE}\cite{europvm,pvmgrace}. 
\par
However, indeed, the execution time required for calculating each
Feynman diagram varies diagram by diagram because diagrams may have different kinds of 
couplings and internal lines.
This small fluctuation causes the load imbalance among parallel
processors and leads to the decrease of the performance.
\par
\begin{figure}
\begin{center}
\rotatebox{270}{\epsfig{file=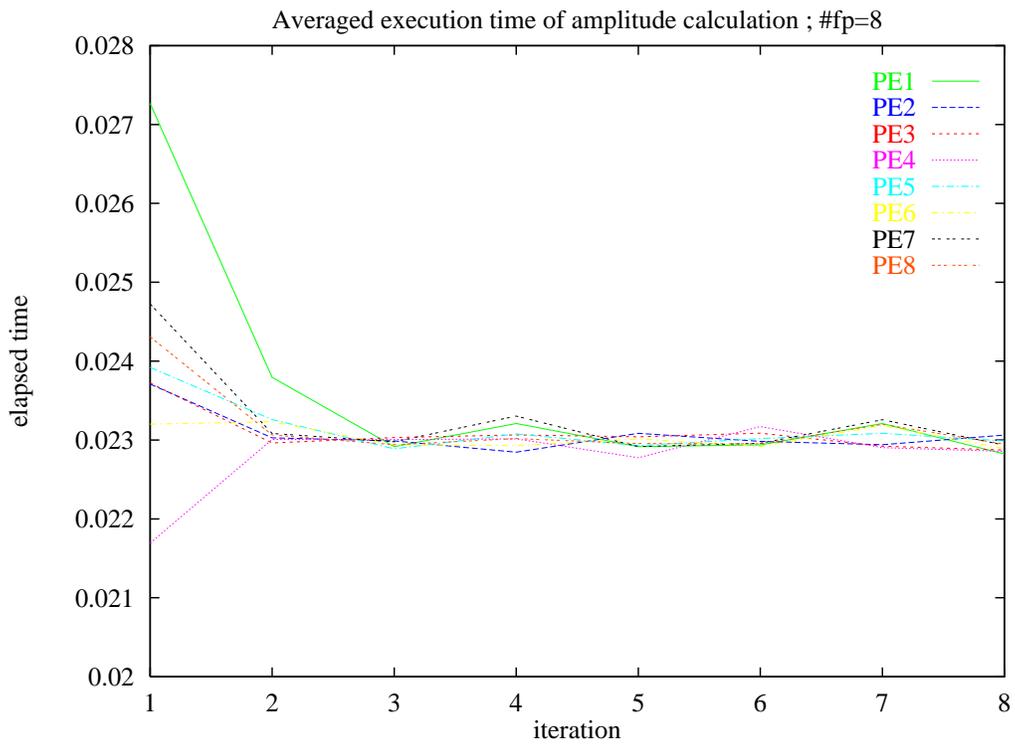,height=14cm}}
\end{center}
\caption{Averaged execution time for calculating scattering
amplitudes in each sampling point in each PE. A vertical axis is averaged
execution time in second. A horizontal axis is the number of iterations.} \label{fig:load-balance}
\end{figure}

We newly add the mechanism of the dynamic load
balancing.  
The execution time of each diagram is
automatically measured in the calculation of the scattering amplitudes.
As the procedures 1. - 5. described in section 2 are
iterated, diagrams are sorted in order of the height of the
load. 
With these rearrangements of diagrams, the load on each parallel
processor becomes well balanced. 
In Fig.~\ref{fig:load-balance}, the way how the fluctuations of the load on each processor
becomes small is shown when the degree of the Function Parallel is
8 and the number of Feynman diagrams is 232.
\par
\section{Performance Measurement}\label{performance}
The performance of the parallel computation is often represented by the
scalability which is 
the ratio of the resultant execution time to the execution time when the
degree is 1.
\par
Table~\ref{tab:scalability} shows the scalability when we use
up to 16 processors. 
As described in section 2, 
the number of parallel processors is given as the
product of the degree of the Data Parallel and that of the Function
Parallel though both can be set independently. 
The degree of the Function Parallel and the degree of the Data
Parallel varies from 1 to 16 so as to keep the product of them
equal to 16. 
\begin{table}
\caption{Reduction rate of the execution time  with hybrid
method. The figures in the columns 
show the behavior of the scalability with fixed degree of the Function
Parallel. The figures in the rows show the behavior of the
scalability with fixed degree of the Data Parallel.}
\label{tab:scalability}
\begin{center}
\begin{tabular}{|l|r|r|r|r|r|} \hline 
Function &\multicolumn{5}{|c|}{Data} \\ \hline
  & 1   & 2   & 4   & 8   & 16   \\ \hline
1 &1.00 &1.97 &3.97 &8.03 &15.82 \\ \hline 
2 &2.02 &3.95 &7.60 &15.56& -     \\ \hline
4 &3.91 &7.57 &14.96& -    & -     \\ \hline
8 &7.15 &13.79& -    & -    & -     \\ \hline
16&12.02& -    & -    & -    & -     \\ \hline
\end{tabular}
\end{center}
\end{table}
\par
The physics process used in the measurement is :
\begin{itemize}
\item{a physics process $e^+e^- \rightarrow
b\bar{b}u\bar{d}\bar{\nu_{\mu}} \mu$, and }
\item{the total number of Feynman diagrams involved in this physics
process is 232 at a tree level with the unitary gauge.}
\end{itemize}
The performance measurement has been performed on AP3000 in Fujitsu Parallel Computing Research Center in Kawasaki, Japan.
The AP3000 system consists of UltraSPARC -II 300MHz processors
connected via AP-Net\footnote{AP-Net is a
two-dimensional torus network.} providing 200MB/s bandwidth per port.
Of data-transfers among processors, the results gathered
in {\tt fpcom} are the biggest and the size of them is about 8.4MB per each 
transfer for the above physics process. In this measurement, the number of
data-transfer is 8 times.
\par
Table ~\ref{tab:scalability} clearly shows that the parallel
computation gives an excellent effect to reduce the practical execution
time of the  
multi-dimensional integration in {\tt GRACE}.

\section{Conclusion}
\noindent
We implemented a hybrid method of the Data Parallel and the Function
Parallel in the multi-dimensional integration of {\tt GRACE}. 
In the parallel computation of Monte Carlo
integration, not only the sampling points but also the calculations of
scattering amplitudes are distributed into parallel processors  .
We used {\tt MPI-1} as Message Passing Library.
In the hybrid method, we can set the degree of the Data Parallel 
and that of the Function Parallel independently. The total number of
parallel processors is given as the product of these two degrees. 
The computing model we used is {\tt SPMD} model.
To reduce the load imbalance among processors due to the small
fluctuations of 
execution time for calculating each Feynman diagram, we   
implemented the mechanism of the dynamic load balancing for distributing
Feynman diagrams to processors. With this mechanism, the effect to the
reduction of 
the elapsed time in {\tt GRACE} has been improved.
\par
The reduction rate of the execution time has been measured on
Fujitsu AP3000 system by using up to 16 processors for the physics
process $e^+e^- \rightarrow b\bar{b}u\bar{d}\bar{\nu_{\mu}} \mu$.
In this study, we found when the program size is not large, the Data Parallel
gives better performance than the Function Parallel.
When, however, the program size is large, it is impossible to run the
program on a single processor. In this case, we have to take the
Function Parallel. Our study tells even the Function Parallel gives
satisfactory results.  
Further, when the network speed is enough high, it is expected that
the performance by the Function Parallel becomes high as the Data
Parallel. 

\section*{Acknowledgments}
\noindent
We wish to thank the members of MINAMI-TATEYA collaboration for
continuous discussions and many kinds of support.
We are also grateful to express our sincere gratitude to
Prof. Y.Shimizu for the valuable suggestions and continuous encouragements. 
Authors appreciate Prof. Y.Watase for the encouragements. 
%
%

%

\begin{thebibliography}{3}
%
\bibitem{GRACE} 
MINAMI-TATEYA group:
{\tt GRACE} manual.
KEK Report 92-19.
%
%
\bibitem{BASES1} 
S.Kawabata:
A new Monte Carlo event generator for High Energy Physics.
Computer Physics Communication {\bf 41} (1986) 127-153.\\
%
\bibitem{BASES2}
S.Kawabata:
A New Version of the Multi-dimensional Integration and Event
Generation Package {\tt BASES/SPRING}.
Computer Physics Communication {\bf 88} (1995) 309-326.
%
\bibitem{parallelbases1} 
T.Ishikawa:
Talk in Future in HEP Computing. KEK Proceedings. Feb (1993) 167-179.
%
\bibitem{parallelbases2} 
T.Ishikawa and S.Kawabata:
Monte Carlo Integration on a parallel computing.
ANNUAL REPORT 1992-1993 Project Directory published by Fujitsu
Parallel Computing Research Facilities, June 20 (1993).
%
\bibitem{europvm}
F.Yuasa et al.:
Running {\tt PVM-GRACE} on Workstation Clusters.
Lecture Notes in Computer Science (1996) 335-338.
%
\bibitem{pvmgrace}
F.Yuasa et al.:
{\tt PVM-GRACE}.
Nuclear Instruments and Methods in Physics Research {\bf A389} (1997)
77-80.
%
\bibitem{mpi}
Message Passing Interface Forum. {\tt MPI}: A message passing interface
standard. Computer Science Dept. Technical Report CS-94-230,
University of Tennessee, Knoxville, TN, April 1994.
%
\end{thebibliography}
%
%
\end{document}